\documentclass[aip,jap,reprint,frontmatterverbose]{revtex4-1}
\usepackage{graphicx}
\usepackage{ulem}
\usepackage{amssymb}
\usepackage{amsmath}
\usepackage{todonotes}
\usepackage{wasysym}
\usepackage{relsize}
\usepackage{bm}
\bibpunct{}{}{,}{s}{}{}
\usepackage{float}
\usepackage{epstopdf}

\begin{document}

\title{Gamow factors and current densities in cold field emission theory: a comparative study}

\author{Debabrata Biswas}\email{dbiswas@barc.gov.in}
\affiliation{
Bhabha Atomic Research Centre,
Mumbai 400 085, INDIA}
\affiliation{
  Homi Bhabha National Institute, Mumbai 400 094, INDIA}

\begin{abstract}
  The factors that contribute to the accuracy of the cold field emission current within
  the contemporary frameworks are investigated. It is found that so long as the net current is
  evaluated using an expression for the local current density obtained by linearizing
  the Gamow factor, the primary source of error is  the
  choice of the energy at which the Taylor expansion
  is done, but not as much on the choice of the method used to
  arrive at the approximate Gamow factor. A suitable choice of linearization energy
  and the implementation of the Kemble correction, allows
  the restriction of errors to below 3\% across a wide range of local fields.
\end{abstract}

\maketitle

\newcommand{\be}{\begin{equation}}
\newcommand{\ee}{\end{equation}}
\newcommand{\bea}{\begin{eqnarray}}
\newcommand{\eea}{\end{eqnarray}}
\newcommand{\Tbar}{{\bar{T}}}
\newcommand{\En}{{\cal E}}
\newcommand{\K}{{\cal K}}
\newcommand{\GC}{{\cal \tt G}}
\newcommand{\Lop}{{\cal L}}
\newcommand{\DB}[1]{\marginpar{\footnotesize DB: #1}}
\newcommand{\q}{\vec{q}}
\newcommand{\kt}{\tilde{k}}
\newcommand{\Lopn}{\tilde{\Lop}}
\newcommand{\noi}{\noindent}
\newcommand{\ovn}{\bar{n}}
\newcommand{\ovx}{\bar{x}}
\newcommand{\ovE}{\bar{E}}
\newcommand{\ovV}{\bar{V}}
\newcommand{\ovU}{\bar{U}}
\newcommand{\ovJ}{\bar{J}}
\newcommand{\calE}{{\cal E}}
\newcommand{\ovphi}{\bar{\phi}}
\newcommand{\zt}{\tilde{z}}
\newcommand{\rt}{\tilde{\rho}}
\newcommand{\tth}{\tilde{\theta}}
\newcommand{\nuv}{{\rm v}}
\newcommand{\ck}{{\cal K}}
\newcommand{\cc}{{\cal C}}
\newcommand{\ca}{{\cal A}}
\newcommand{\cb}{{\cal B}}
\newcommand{\cg}{{\cal G}}
\newcommand{\ce}{{\cal E}}
\newcommand{\cn}{{\cal N}}
\newcommand{\ct}{{\cal T}}
\newcommand{\cp}{{\cal P}}
\newcommand{\fn}{{\small {\rm  FN}}}
\newcommand\norm[1]{\left\lVert#1\right\rVert}
\newcommand{\Afn}{A_{\text{FN}}}
\newcommand{\Bfn}{B_{\text{FN}}}
\newcommand{\mg}{\text{MG}}
\newcommand{\sh}{\text{SF}}



\section{Introduction}
\label{sec:intro}

The evaluation of the cold field emission current from a metallic surface has been the
subject of interest for almost a century starting with the works of Fowler-Nordheim (FN),
Murphy-Good (MG) and several others in the past decades\cite{FN,Nordheim,murphy,jensen2003,
forbes2006,FD2007,DF2008,jensen_book,jensen_rTF,db_dist,db_rr_2019,rr_db_2021,db_rr_2021,ludwick}.
The analytical expression for the current density is based on the free-electron
model of metals, the fermionic nature of electrons, the WKB approximation for the
tunneling transmission coefficient\cite{TC} and finally an integration over the electron states
using a linearization of the Gamow factor. Having a ready-to-use  expression
is helpful since it speeds up the numerical calculation of the net emission current by
integrating over the surface. It also allows for an approximate analytical expression for
the net emission current using (in several cases) a knowledge of the local field variation
around the emitter apex\cite{db_ultram,physE}. Finally, an analytical expression
for the distribution of emitted electrons\cite{db_dist} also leads to a better
and faster modelling of the emission process
in Particle-In-Cell codes\cite{sg_rk_gs_db_2021,rk_gs_db_2021,db_rk_gs_2021}.

The continued theoretical interest in the subject stems from the fact that even
for cold field emission, there remains sufficient scope for improving the modelling
of  emitters based on its validation with experimental results. For instance,
the Fowler-Nordheim\cite{FN} expression for current density
(see for instance Eq.~(6) of Ref. [\onlinecite{forbes2019jap}])
that uses the exact triangular (ET) barrier potential,
is almost certainly invalid for its neglect of image charge
contributions included in the Murphy and Good form\cite{murphy},
a point emphasized by Forbes\cite{forbes2019jap}. A manifestation of
its shortcoming is that emitter characteristics, such as the field enhancement factor
or the emission area, inferred from the experimental data are highly
inconsistent with the physical dimensions of the emitter\cite{db_rk_2019,db_universal}.
The Murphy-Good current density on the other hand,
seems to capture the essential physics \cite{forbes2019jap}
but is inadequate  for nano-tipped
emitters having apex radius of curvature less than 100nm\cite{db_rr_2019,db_rk_2019}.
Thus, incorporation of curvature effects is an area of concern and
future efforts in this direction may be aimed at further
improvement of accuracy.

The focus in this work is on emitters where curvature effects are negligible
and hence they can be adequately described within the standard MG approach.
The errors in the Murphy-Good current density (MGCD) in
the high field regime\cite{jensen_rTF,db_rr_2021} is a motivation for the present study. 
There are three basic ingredients that go into the derivation of MGCD
that need to be looked at afresh. The first of these centres around the Gamow factor,

\be
G(\ce) = g \int_{s_1}^{s_2} \sqrt{V_T(s) - \ce}~ ds
\ee

\noi
where $V_T$ is the tunneling potential energy, $s_1$ and $s_2$ are the turning points
determined using $V_T(s) - \ce = 0$, $g = \sqrt{8m}/\hbar$, $m$ is the mass of the
electron and $\ce$ is the normal-energy of the electron incident on the barrier.
In the standard Murphy-Good approach, the Gamow factor $G$ for the
image charge modified potential energy (the Schottky-Nordheim or SN barrier; see Eq.~(\ref{eq:SN})),
is expressed as a product of the Gamow factor for the exact triangular barrier
$G_{\text{ET}} = (2/3) g \varphi^{3/2}/(qE_l)$,
and a barrier form correction factor (BFCF). For the SN barrier, the
BFCF is the WKB integral

\be
\begin{split}
\nu(y)  = & \frac{3}{2} \int_{\xi_1}^{\xi_2} d\xi~
\left( 1 - \xi + \frac{y^2}{4\xi} \right)^{1/2}  \label{eq:nu}
\end{split}
\ee

\noi
where $\xi_1$ and $\xi_2$ are the roots of $1 - \xi + y^2/(4\xi)$ with $y = 2\sqrt{B qE_l}/\varphi$.
Here $B =  q^2/(16\pi\epsilon_0)$, $E_l$ is the local field, $\varphi =  \ce_F + \phi - \ce$,
$q$ is the magnitude of the electronic charge, $\ce_F$ is the Fermi energy, and $\phi$ is the workfunction. 
The BFCF $\nu(y)$ due to the image charge can also be expressed in terms of
complete elliptic integrals (see Eq.~(16) of Murphy and Good\cite{murphy}).
Importantly, for this communication, $\nu(y)$ has a convenient algebraic approximation
$\nu(y) \approx 1 - y^2 + (y^2/3)\ln(y)$
due to Forbes\cite{forbes2006} which allows the MGCD to be readily used to evaluate the field emission
current density. In the following, we shall refer to the Gamow-factor
evaluated using the Forbes approximation as the
MG method of evaluating the Gamow factor and refer to this as $G_\mg$.
Obviously, $G_\mg$ provides  a useful approximate value for the `exact' Gamow factor, $G$,
which can be obtained by numerically integrating Eq.~(\ref{eq:nu}) and
multiplying this by $G_{\text{ET}}$. The `exact-WKB' method that we shall
use in the following sections as a benchmark, uses the `exact' Gamow factor. 

An alternate, though related, approach to the evaluation of the Gamow factor, is the so-called
shape-factor (SF) method due to Jensen\cite{jensen2012_sf,jensen2017,jensen_rTF}, which
recasts the Gamow factor as $G = 2 L(y) \kappa(y) \sigma(y)$, a product form
(see Eqns.~\ref{eq:GSF}-\ref{eq:SFS}) applicable to all barriers,
with individual terms dependent on the shape of the barrier.
The so-called shape-factor term in the product, $\sigma(y)$, is an
integral (see Eq.~(\ref{eq:SFS})) that still needs to be evaluated.
Note that that $G = G_{\text{ET}}~ \nu(y)$ is exactly equivalent to $G = 2 L(y) \kappa(y) \sigma(y)$.
Both forms should lead to the exact Gamow factor if $\nu(y)$ in Eq.~(\ref{eq:nu}) and
$\sigma(y)$ in Eq.~(\ref{eq:SFS}) are computed accurately. 

For the SF method to be readily used, an algebraic approximation
of the shape-factor integral is required just like the Forbes approximation for $\nu(y)$.
Useful approximations for the shape-factor  $\sigma(y)$ for the SN barrier exist,
expressed as a second or fourth degree polynomial in $(1-y)/(1+y)$,
with the coefficients determined from a fit. We shall refer to Gamow factor evaluated thus as $G_{\text{SF}}$.
Both $G_\mg$ and $G_{\text{SF}}$ represent basic approximations
that can be used to arrive at expressions for the field emission current density
and the present comparative study involves these approximate forms of the Gamow factor.
As we shall see, $G_{\text{SF}}$ leads to a more accurate evaluation of the
transmission coefficient compared to $G_\mg$, for both the second and fourth degree
approximate polynomial forms of the shape factor.

The second of the three ingredients necessary for an analytical expression for
the current density, involves the point on the energy scale at which a linearization of the
Gamow factor should be made. In the standard MG approach, the Fermi energy has been the
point of linearization. While this may be adequate for cold field emission at low
to moderate fields, it is clear that this would lead to large errors at higher fields or in the
thermal-field region. The emerging point of view is
that the location of the peak in the normal energy distribution provides a suitable energy
for linearization. The shifted point of linearization, together with the use of
the shape factor approximation, results in a more accurate determination of the
current density in case of thermal-field emission\cite{jensen_rTF}.
In case of cold field emission, the shape factor method (even for second degree approximation)
yields a somewhat more involved expression for the current density 
compared to the MGCD as we shall see in section \ref{sec:CD}. Whether this translates to a more
accurate expression for the current density will be a subject of investigation.

The third ingredient concerns the use of $e^{-G}$ for the transmission
coefficient. While this is adequate at energies where the barrier is strong, it
contributes to larger errors at energies close to the top of the barrier. At higher field strengths,
the SN-barrier peak comes closer to the Fermi energy while the peak of the
normal energy distribution shifts away from the Fermi energy. Thus, the second
approximation (linearization at Fermi energy) as well as the third approximation (use of $e^{-G}$)
contribute to the errors at higher fields.
An alternate and better approximation near the barrier-top is the so-called Kemble
form which uses $(1 + e^G)^{-1}$ to approximate the transmission
coefficient within the WKB method.

While there has been a need to correct the errors in the prediction of the MGCD
in the high field regime, it is of interest to know whether the
shape-factor method 
has an equally important role to play in improving the accuracy, as the point
of linearization along the energy axis and the
Kemble form of transmission coefficient.
The interest in such a question is manifold apart from the purely academic one.
To begin with, if the simplicity of the MGCD can be
retained without compromising with the accuracy by merely shifting the
point of linearization and introducing a correction term to account
for the Kemble form, such an approach might be worthwhile. Besides, 
the entire paraphernalia of curvature-corrections to the current density\cite{db_rr_2021}
has been computed as an extension of MGCD, and
it would be helpful to know if the errors can be reduced by merely
switching the point of linearization and introducing a correction if necessary.
This is also true for the existing
expressions for the net field emission current from locally parabolic tips and
the electron distributions\cite{db_dist}. We shall thus compare the field emission current density
and net emission current using $G_\mg$ and $G_{\text{SF}}$, and study the relative
importance of the point of linearization, the two approximate methods of
evaluating the Gamow factor, and the use of Kemble form in the context of cold field emission.
To keep matters simple, we shall assume that the emitters have
tip-radius large enough to ignore curvature corrections.

Before embarking on this comparative study, it is important to decide on
the benchmark to be used. Since the WKB method is central to the field
emission formalism, it is essential to use an exact numerical evaluation of the Gamow factor,
and the Kemble form of the transmission coefficient to 
compute the current-density by integrating over
the electron states. We shall refer to this as the exact-WKB method, the word `exact'
referring to the use of the exact Gamow factor and numerical integration over the energy states
but not to the transmission coefficient.
This approach provides a natural benchmark for comparing
other approximate results which invoke an approximation to the
Gamow factor and its subsequent linearization in order to carry out the energy
integration. It is important to note that the exact current density,
which can be computed for instance by using
the transfer matrix approach\cite{DB_VK},
may differ substantially from the benchmark itself\cite{mayer2011,db_rr_2021}, depending on the ratio
of the  Fermi energy and the workfunction\cite{db_rr_2021}.

In section \ref{sec:TC}, we shall first compare the  transmission coefficient
for different ranges of energy using $G_\mg$ and $G_{\text{SF}}$, and also look at
the net emission current without resorting
to linearization. Section \ref{sec:CD} deals with analytical expressions for
current density using an approximate Kemble form and a linearization of the Gamow factors
$G_\mg$ and $G_{\text{SF}}$ at an arbitrary energy $\ce_m$ at or below the Fermi energy.
These are then used to compare the net emission current in the linearized
framework with the benchmark. Our conclusions
and discussions form the final section.

\section{Comparison of the transmission coefficient}
\label{sec:TC}

The central object in field emission is the tunneling transmission coefficient, $T$.
The WKB-method provides a handy method for determining $T(\ce)$ for a
particle with incident normal-energy $\ce$. It may be expressed as\cite{kemble,forbes2008pre} 

\bea
T(\ce) & \approx  & \frac{1}{1 + e^{G(\ce)}} \label{eq:Kemble} \\
G(\ce) & = & g \int_{s_1}^{s_2} \sqrt{V_T(s) - \calE}~ds
\eea

\noi
where $g = \sqrt{8m}/\hbar \simeq 10.246 \text{(eV)}^{-1/2} \text{(nm)}^{-1}$
while $s_1$,$s_2$ are the zeroes of the integrand. The tunneling potential energy

\be
V_T(s) = \ce_F + \phi  - qE_ls -  \frac{B}{s} \label{eq:SN}
\ee

\noi
where $q$ is magnitude of the electronic charge, $E_l$ is the local
field a point on the emitter-surface, $s$ is the normal distance from the
point, $B =  q^2/(16\pi\epsilon_0)$ while $\ce_F$ and $\phi$ are the
Fermi energy and workfunction respectively. To simplify matters, we have not
included any curvature correction to the tunneling potential energy. Note
that the exact-triangular potential energy can be obtained by neglecting the image
charge contribution $B/s$ in Eq.~(\ref{eq:SN}).

In the Murphy-Good (MG) approach, the Gamow factor $G(\ce)$ is expressed
as \cite{murphy}

\be
G(\ce) = g \frac{2}{3} \frac{\varphi^{3/2}}{qE_l} \nu(y) = G_{\text{ET}}~ \nu(y)  \label{eq:GMG}
\ee

\noi
where $\varphi =  \ce_F + \phi - \ce$, $y = 2\sqrt{B q E_l}/\varphi$ and
$E_l$ is the local electric field. The BFCF or the image-charge correction factor,
$\nu(y)$ is well approximated
by\cite{forbes2006}

\be
\nu(y) \approx 1 - y^2  + \frac{y^2}{3} \ln(y).  \label{eq:forbesnu}
\ee

\noi
Eq.~(\ref{eq:GMG}) together with Eq.~(\ref{eq:forbesnu}) gives an approximate
expression for the Gamow factor and is referred to as $G_\mg$.

In the more recent shape-factor (SF) approach, the Gamow factor is
expressed as\cite{jensen2012_sf,jensen_rTF}

\be
G(\ce) = 2 \sigma(\ce) \kappa(\ce) L(\ce).  \label{eq:GSF}
\ee

\noi
For the Schottky-Nordheim barrier,

\bea
L(\ce) & = & \frac{1}{E_l} \sqrt{\varphi^2 - 4BE_l} = \frac{\varphi}{E_l}(1 - y^2)^{1/2} \label{eq:SFL} \\
\kappa(\ce) & = & \frac{g}{2} (\varphi - \sqrt{4BE_l})^{1/2} = \frac{g \sqrt{\varphi}}{2} (1 - y)^{1/2} \label{eq:SFK} \\
\sigma(y(\ce)) & = & \frac{\sqrt{2}}{4} (1 + y)^{1/2} \int_{-1}^{1} ~ds~\left[ \frac{1 - s^2}{1 + s\sqrt{1 - y^2}} \right]^{1/2} .  \label{eq:SFS}
\eea

\noi
Useful approximate forms for the shape factor $\sigma(y)$ exist, expressed
as\cite{jensen_rTF}

\be
\sigma(y) = \sum_{j=0}^{n} C_j \left(\frac{1 - y}{1 + y} \right)^{j}. \label{eq:SFS_approx}
\ee

We shall, for the most part,  use the one with $n = 2$ with $C_0 = 0.785398$, $C_1 = -0.092385$
and $C_2 = -0.026346$ in order to obtain a 
manageable expression for the linearized current density in section \ref{sec:CD}.
Eqn.~(\ref{eq:GSF}) together with Eqns.~(\ref{eq:SFL}),(\ref{eq:SFK}) and (\ref{eq:SFS_approx})
gives an approximate expression for the Gamow factor in the SF approach and is referred to as $G_{\text{SF}}$. 

These approximate expressions for the Gamow factor can be used to arrive at expressions
for the current density

\be
J = \frac{2mq}{(2\pi)^2 \hbar^3} \int_{0}^{\calE_F} T({\ce})~ (\ce_F - \ce)~ d{\ce}   \label{eq:Jbasic}
\ee

\noi
which can finally be integrated over the surface to arrive at the net emission current.

It is instructive to compare the transmission coefficient and net emission current obtained using
the two approaches before proceeding with the linearization of the Gamow factor to obtain
an analytical expression for the current density.

\begin{figure}[thb]
\vskip -0.5 cm
\hspace*{-01.05cm}\includegraphics[width=0.6\textwidth]{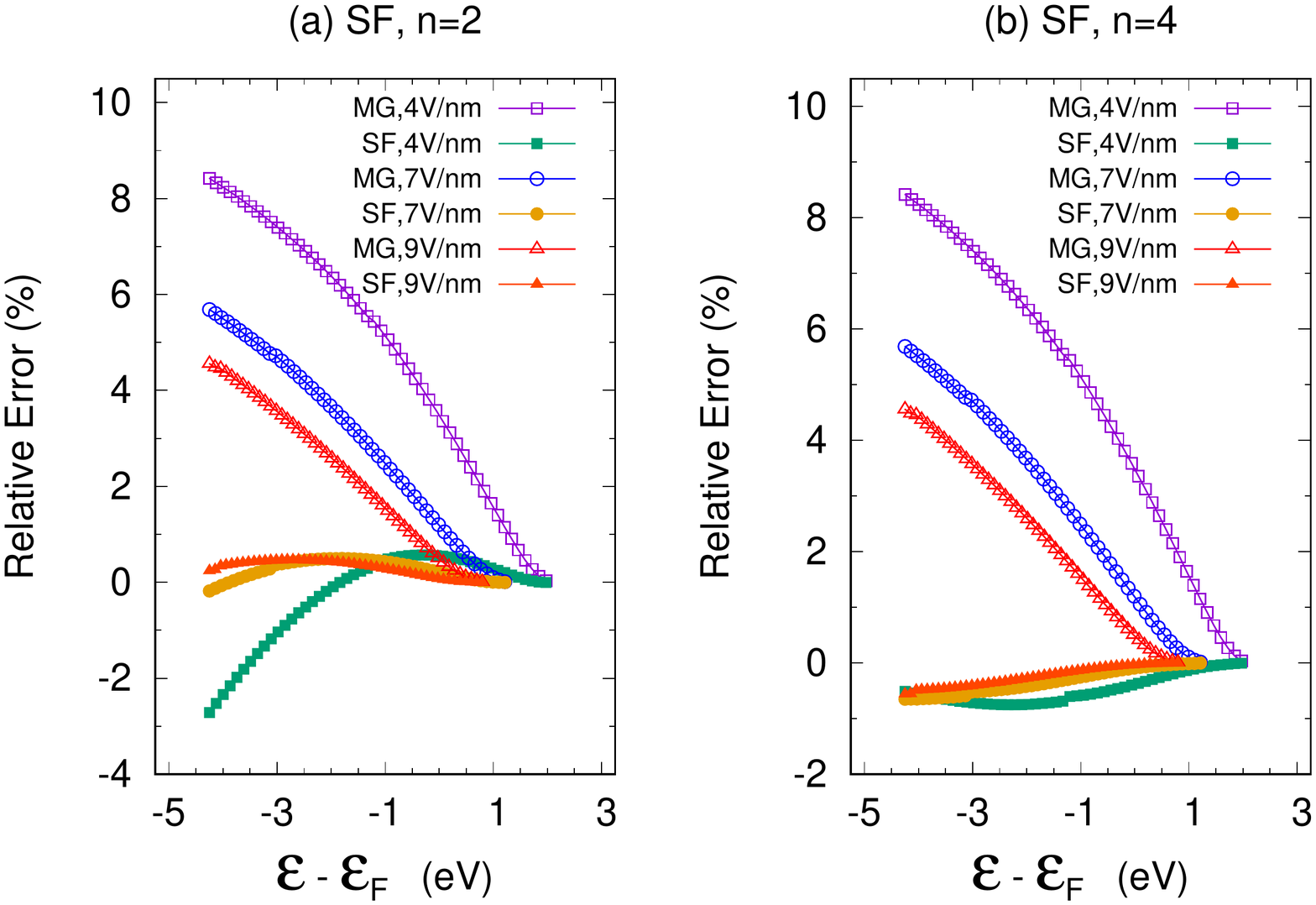}
\vskip -0.75 cm
\caption{The relative error in transmission coefficient $T(\ce)$ with respect to the exact-WKB result. The maximum value of energy $\ce$ for each applied field corresponds to the top of the barrier. MG refers to Murphy-Good while SF refers to the Shape-Factor method. In (a) the second degree polynomial form of the shape factor is used with $C_0 = 0.785398$, $C_1 = -0.092385$ and $C_2 = -0.026346$. In (b) the fourth degree polynomial form is used with $C_0 = 0.785398, C_1 = -0.0961, C_2 = -0.029092, C_3 = 0.034482$ and $C_4 = -0.027987$. Also marked are the local fields in V/nm. At higher fields, the barrier comes closer
to the Fermi energy.}
\label{fig:errorTC}
\end{figure}

\begin{figure}[thb]
\vskip -0.75 cm
\hspace*{-01.05cm}\includegraphics[width=0.6\textwidth]{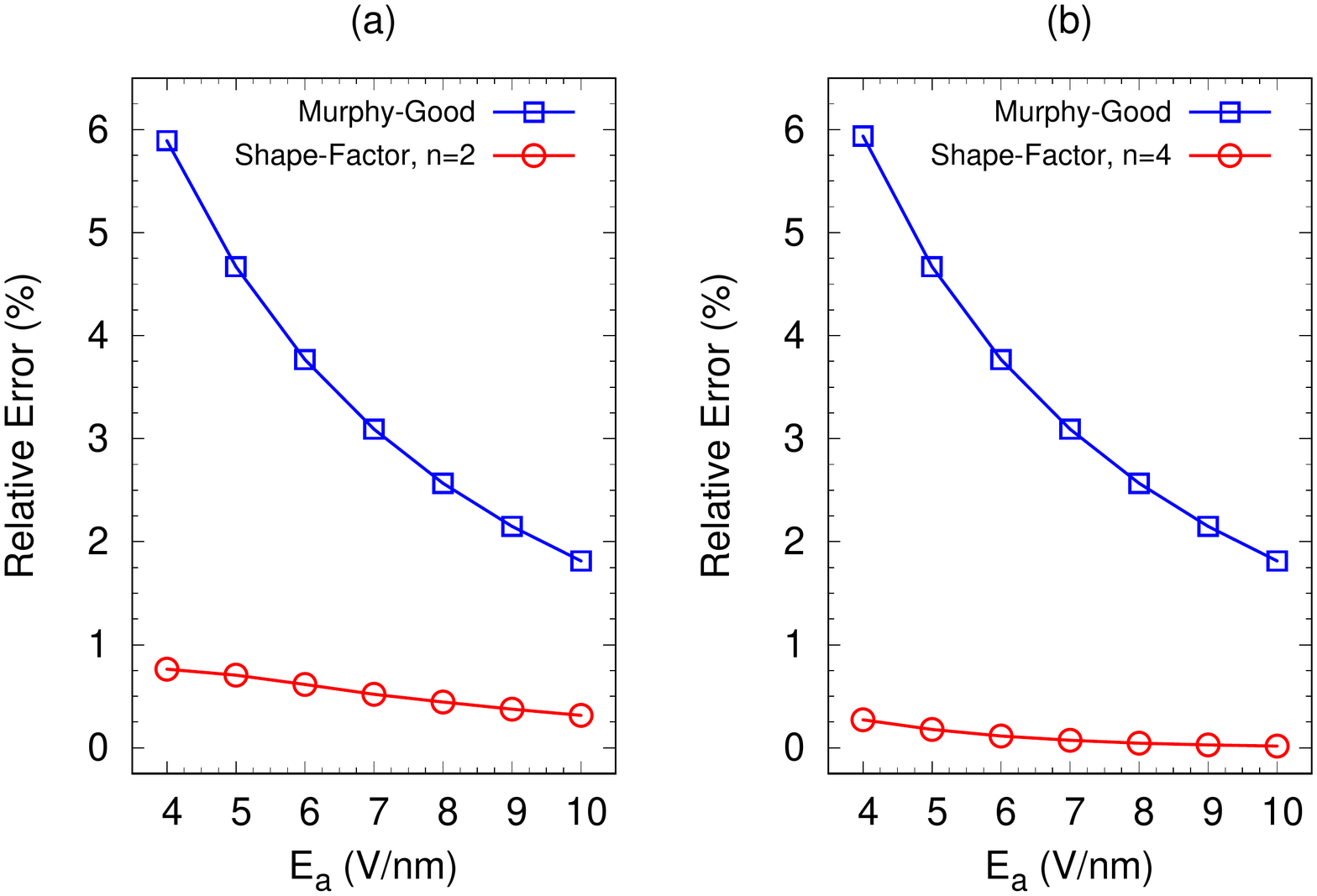}
\vskip -0.75 cm
\caption{The relative error in net-emission current with respect to the exact-WKB result.
  Both SF and MG methods use the Kemble form and numerical integration over energy. Both under-predict the current.
  In (a) the second degree polynomial form approximation of the shape factor is used while in (b) the
fourth degree polynomial form is used. The coefficients $C_i$ are as in Fig.~\ref{fig:errorTC}.}
\label{fig:errorI}
\end{figure}

Figure~(\ref{fig:errorTC}) shows the error in transmission coefficient as a function of energy for
3 different applied fields, $E_l = 4, 7$ and $9$V/nm, for a material with $\phi = 4.5$eV and
$\ce_F = 8.5$eV. The label MG refers to the transmission coefficient evaluated using
$T_\mg(\ce) \approx 1/(1 + e^{G_\mg(\ce)})$ with $G_\mg$ evaluated using Eq.~(\ref{eq:GMG})
and Eq.~(\ref{eq:forbesnu}).
Similarly, SF refers to the transmission coefficient evaluated as $T_{\text{SF}} \approx
1/(1 + e^{G_\text{SF}})$
with ${G_\text{SF}}$ evaluated using Eqns.~(\ref{eq:GSF}), (\ref{eq:SFL}), (\ref{eq:SFK}) and
(\ref{eq:SFS_approx}). Clearly, the
shape factor approximation of the Gamow factor provides much better results for both $n=2$ and $n=4$
at all energies while the MG transmission coefficient
scores well at higher energies. Note that the maximum value of energy for each applied field
corresponds to the top of the barrier. It is thus higher for lower field strengths.

Note that insofar as the net emission current is concerned, the electrons that
contribute at a lower apex field strength (such as 4V/nm) have energies in the vicinity of $\calE_F$
so that the larger errors away from $\calE_F$ in Fig.~\ref{fig:errorTC} (for MG and n=2 SF) do not have
an important bearing on errors in the the emission current for both MG and SF approximations. 
A plot of the net-emission current using these transmission coefficient confirms this
observation. In Fig.~(\ref{fig:errorI}), we consider a hemiellipsoidal emitter
in a parallel-plate configuration with $h/R_a = 300$ and $R_a = 10\mu$m. The net emission
current is evaluated by integrating the current density obtained using Eq.~(\ref{eq:Jbasic})
over the surface using the local cosine law of field variation $E_l = E_a \cos\tth$ where
$\cos \tth = (z/h)/\sqrt{(z/h)^2 + (\rho/R_a)^2}$. The label `Murphy-Good' refers to the current
obtained using $T_\mg$ in Eq.~(\ref{eq:Jbasic}) while `Shape-Factor' refers to the use of
$T_{\text{SF}}$ in Eq.~(\ref{eq:Jbasic}). In both cases, the energy integration is
performed numerically. The errors computed are relative to the exact-WKB method
where the Gamow factor is obtained numerically. Clearly $G_{\text{SF}}$ gives better results
for both $n=2$ and $n=4$ compared to $G_\mg$ when the current density is obtained by numerically integrating over
the electron energy states.

\section{Linearization and the analytical current density}
\label{sec:CD}

The linearization of the Gamow factor allows us to perform the energy integration
in Eq.~(\ref{eq:Jbasic}) analytically. If the Taylor expansion is done at $\ce = \ce_F$, the errors
are larger at higher field strengths since the peak of the normal energy distribution
lies below the Fermi energy
for cold field emission and moves further away as the local electric field is
increased. A suitable alternate energy value may be chosen in one of several ways.
The one suggested for thermal-field emission  corresponds to the
peak (or maxima) of the normal energy distribution. We shall thus compare the
two expressions for the linearized current densities and the net emission current
obtained using these.

Eq.~(\ref{eq:Jbasic}) for the current density can be written approximately as,

\bea
J & = & \frac{2mq}{(2\pi)^2 \hbar^3} \int_{0}^{\calE_F} (\ce_F - \ce)\frac{1}{1 + e^{G(\ce)}}~d{\ce}  \\
& \approx & \frac{2mq}{(2\pi)^2 \hbar^3} \int_{0}^{\calE_F} (\ce_F - \ce) e^{-G(\ce)} \left[ 1 - e^{-G(\ce)} + \ldots \right]~d{\ce}   \nonumber
\eea

\noi
The integration can now be carried out easily.
Since the algebra is quite straightforward, we shall merely state the final result. In the
MG case, a linearization at $\ce = \ce_m$ leads to

\bea
J_{\mg}^m & \approx & \Afn  \frac{1}{\varphi_m}\frac{E_l^2}{t_m^2} e^{-\cb_\mg} \left( 1 - \frac{e^{-\cb_\mg}}{4} \right) \label{eq:MGshift} \\
\cb_\mg & = & \Bfn \varphi_m^{3/2} \frac{\nu_m}{E_l} + \frac{t_m}{d_m}(\ce_F - \ce_m)    \label{eq:BFNshift}
\eea

\noi
where $A_\fn~\simeq~1.541434~{\rm \mu A~eV~V}^{-2}$,
$B_\fn~\simeq 6.830890~{\rm eV}^{-3/2}~{\rm V~nm}^{-1}$ are the usual Fowler-Nordheim
constants, $g =  \sqrt{8m}/\hbar \simeq 10.246 \text{(eV)}^{-1/2} \text{(nm)}^{-1}$,
$E_l$ is the local field, while $\varphi_m  =  \ce_F + \phi - \ce_m$, $d_m^{-1}  = g \frac{\varphi_m^{1/2}}{E_l}$,
$y_m  =  c_S \sqrt{E_l}/\varphi_m$ with $c_S = 1.199985 {\rm eV}~{\rm V/nm}^{-1/2}$ and

\bea
\nu_m & = & 1 - y_m^2 + \frac{y_m^2}{3} \ln y_m \\
t_m & = & 1 + \frac{y_m^2}{9} - \frac{y_m^2}{9}\ln y_m .
\eea

\noi
Note that the value of $\ce_m$ has not been specified so far and hence Eq.~(\ref{eq:MGshift})
applies to an arbitrary point of linearization between $0$ and $\ce_F$. Recall that
the standard MG approach uses $\ce_m = \ce_F$. Also, the
factor $1 - e^{-\cb_\mg}/4$ in Eq.~(\ref{eq:MGshift}) is a first correction arising from the use of the
Kemble form of transmission coefficient. Neglecting the correction factor,
$1 - e^{-\cb_\mg}/4$, would amount to using $T(\ce) = e^{-G}$ thereby reducing Eq.~(\ref{eq:MGshift})
at $\ce_m = \ce_F$ to the standard MGCD, $J_\mg$. $J_\mg^m$ may thus be referred to as MG-like
current density to emphasize  (a) that $\ce_m$ may be different from $\ce_F$ and
(b) the inclusion of the correction term in $J_\mg^m$.

A similar expression can be obtained using the shape-factor method.
We shall restrict ourselves to the second degree polynomial approximation
in order to obtain a manageable expression for the current density.
The use of $G_{\text{SF}}$ with $n=2$ and its linearization  at $\ce = \ce_m$ leads to a form
for the current density that can be expressed as

\bea
J_{\sh}^m & \approx & \Afn  \frac{1}{\varphi_m}\frac{E_l^2}{\ct_m^2} e^{-\cb_\sh} \left( 1 - \frac{e^{-\cb_\sh}}{4} \right)  \label{eq:SFshift} \\
\cb_\sh & = & \Bfn \varphi_m^{3/2} \frac{\cn_m}{E_l} + \frac{\ct_m}{d_m} (\ce_F - \ce_m) 
\eea

\noi
where

\bea
\cn_m & = & \frac{3}{2} \left[ C_0 y_1y_2^{1/2} + C_1 \frac{y_1^2}{y_2^{1/2}} + C_2 \frac{y_1^3}{y_2^{3/2}}\right] \\
\cp_m & = & \frac{C_0}{2} \frac{y_3}{y_2^{1/2}} + \frac{C_1}{2} \frac{y_4 y_1}{y_2^{3/2}} + \frac{3}{2}C_2 \frac{y_5 y_1}{y_2^{5/2}} \\
 \ct_m & = & \cn_m + y_m \cp_m  
\eea

\noi
with $y_1  = 1 - y_m$, $y_2  =  1 + y_m$, $y_3  =  1 + 3y_m$, $y_4  =  5 + 3y_m$ and $y_5  =  3 + y_m$. The
coefficients  $C_0 = 0.785398$, $C_1 = -0.092385$ and $C_2 = -0.026346$.
As in the MG case, the correction factor $(1 - e^{-\cb_\sh}/4)$  in Eq.~(\ref{eq:SFshift})
accounts for, to a first approximation, the use of the Kemble form of the transmission coefficient.

We are now in a position to compare the relative importance of   the
approximate forms of the Gamow factor, the energy at which they are linearized
and the use of Kemble correction to the transmission coefficient.
For both current densities, the superscript $m$ refers to the point of
linearization.

\begin{figure}[thb]
\vskip -0.5 cm
\hspace*{-01.05cm}\includegraphics[width=0.6\textwidth]{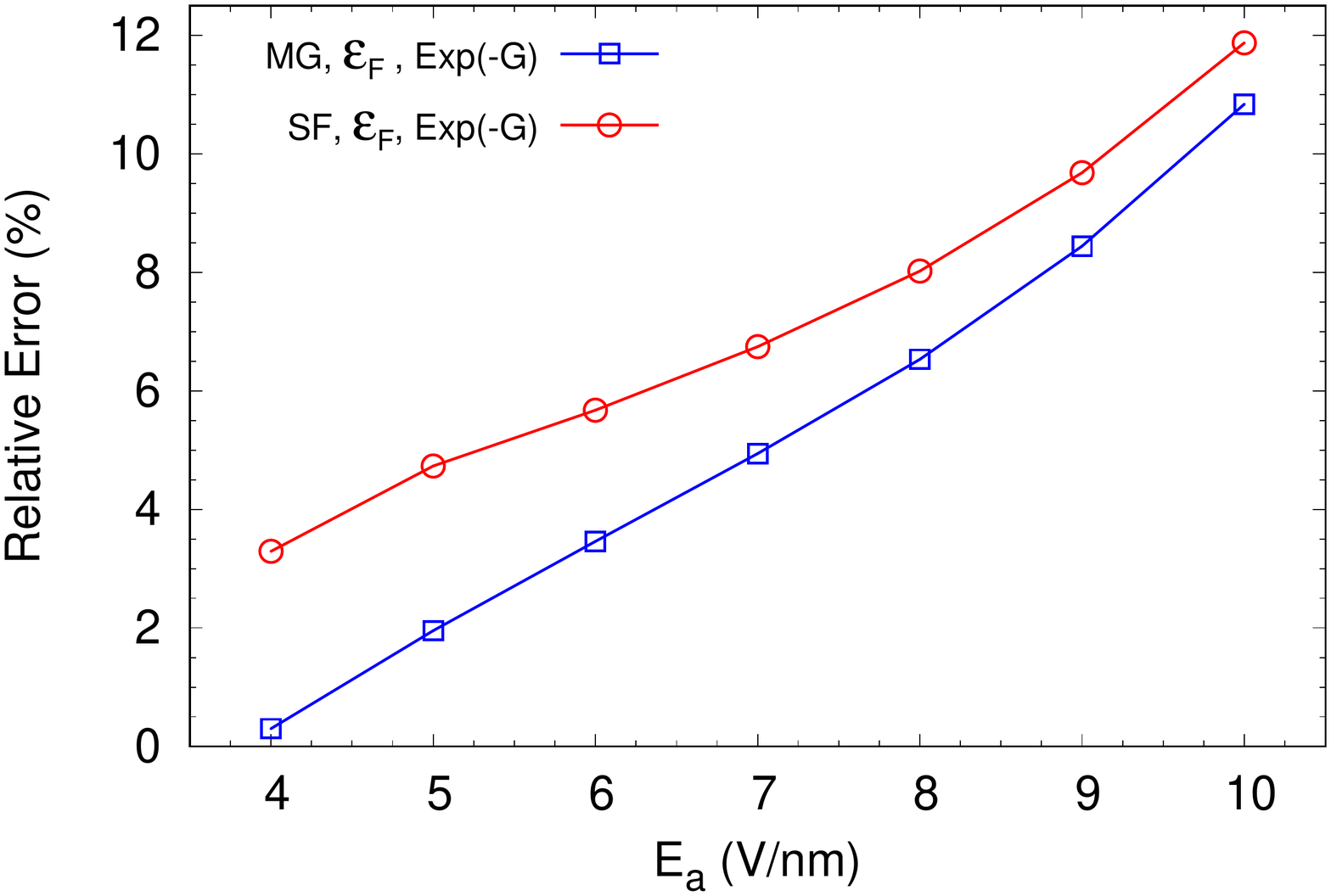}
\vskip -0.75 cm
\caption{The absolute relative error in net-emission current with respect to the exact-WKB result
  calculated using $J_{\mg}^m$ and $J_{\sh}^m$ but without the respective correction factors. Here
  $\ce_m = \ce_F$, the workfunction $\phi = 4.5$eV while $\ce_F = 8.5$eV. Both over-predict the
net current except at $E_a = 4$V/nm for the MG case.}
\label{fig:errorIEf}
\end{figure}

\begin{figure}[thb]
\vskip -0.5 cm
\hspace*{-01.05cm}\includegraphics[width=0.6\textwidth]{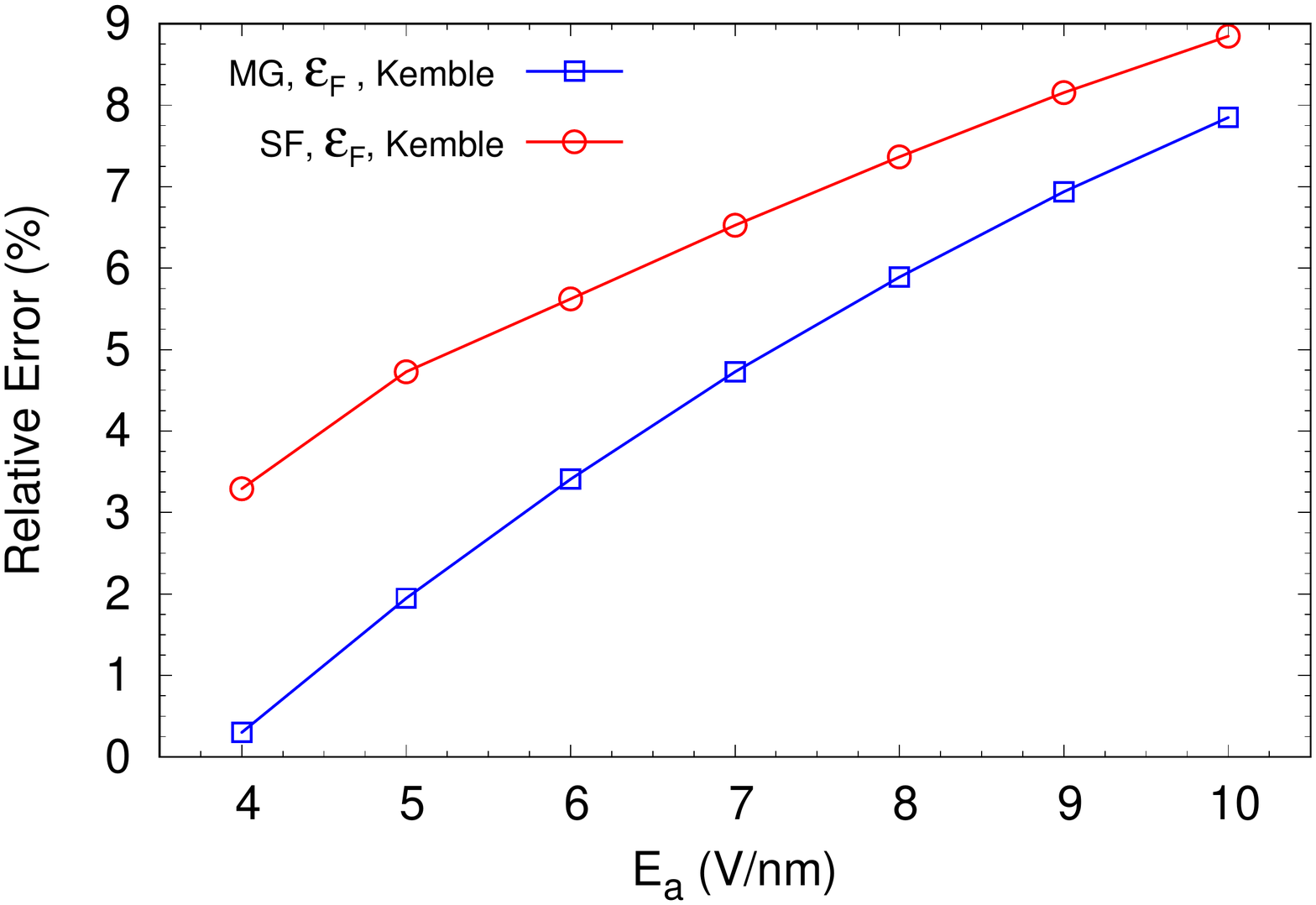}
\vskip -0.75 cm
\caption{The absolute relative error as in Fig.~\ref{fig:errorIEf} using the correction factor in 
  $J_{\mg}^m$ and $J_{\sh}^m$ due to the Kemble form of the transmission coefficient.
The error reduces at higher field strengths but remains unaltered at lower values of $E_a$.}
\label{fig:errorIEfK}
\end{figure}

We shall first compare the net emission current using a linearization at $\ce_F$ by setting
$\ce_m = \ce_F$ for a hemi-ellipsoidal emitter with $h/R_a = 300$ and $R_a = 10\mu$m.
The errors as before are computed with respect to the exact-WKB result which
acts as a natural benchmark. The plot labels refer to the Gamow factor approximation
used (MG vs SF), the value of $\ce_m$ and the approximation used for the transmission
coefficient. Figure \ref{fig:errorIEf} shows the errors in $J_{\mg}^m$
and $J_{\sh}^m$, without the respective correction factors (i.e. using $e^{-G}$
for the transmission coefficient; thus $J_\mg^m$ is $J_\mg$ in this case),
plotted against the apex field $E_a$.
Clearly linearization at $\ce_m = \ce_F$ produces large errors at high fields for both MG and SF.
Surprisingly, it affects the shape factor method more. Note that both MG and SF over-predict
the net current. Fig.~\ref{fig:errorIEfK} shows a similar comparison with the correction
factors in place (i.e. Kemble form).  The errors reduce at higher fields but MG
continues to perform better.

\begin{figure}[thb]
\vskip -0.5 cm
\hspace*{-01.05cm}\includegraphics[width=0.6\textwidth]{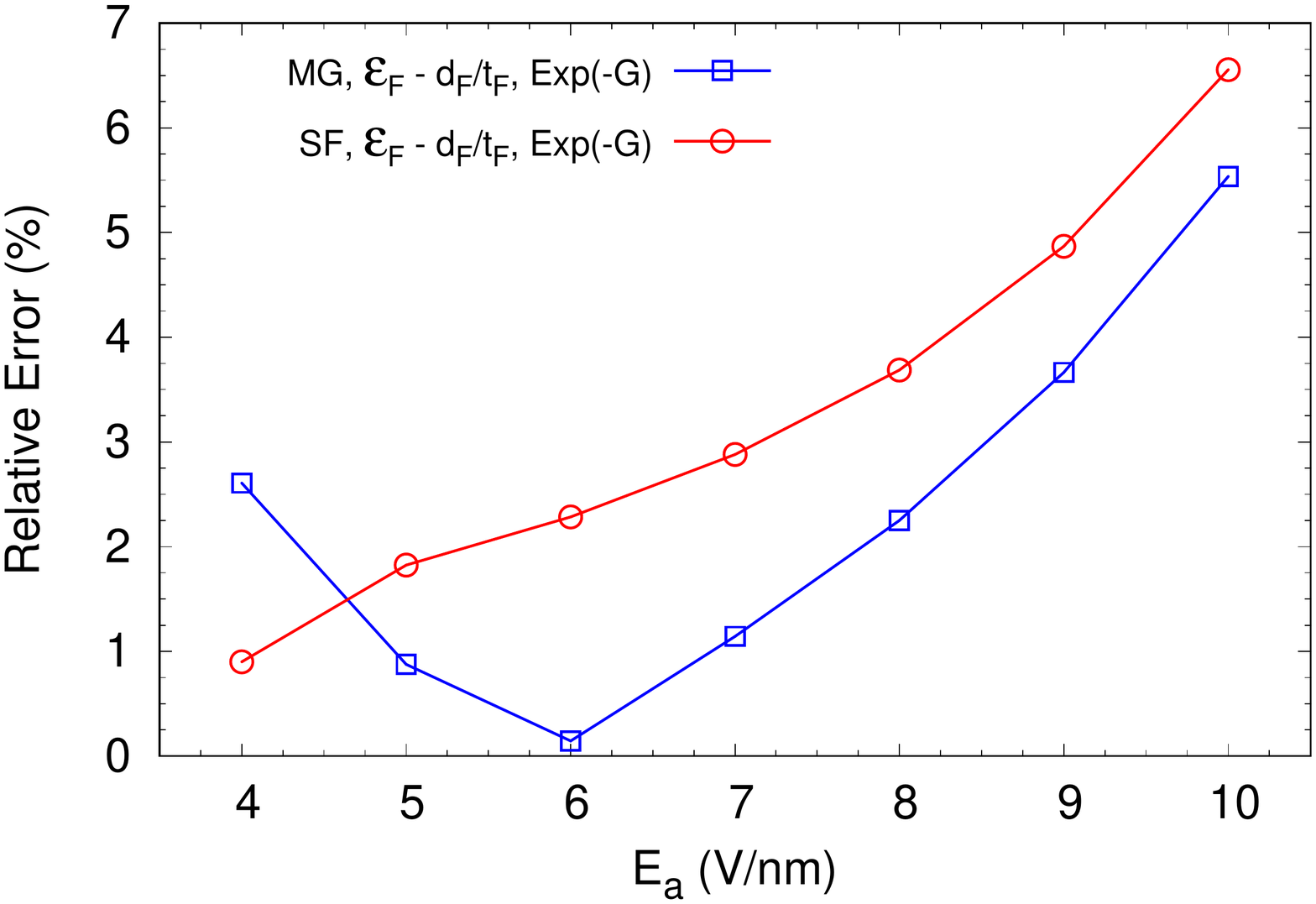}
\vskip -0.75 cm
\caption{The absolute value of the relative error in net-emission current
  with respect to the exact-WKB result
  calculated using $J_{\mg}^m$ and $J_{\sh}^m$ with $\ce_m = \ce_F - d_F/t_F$.
  The workfunction $\phi = 4.5$eV while $\ce_F = 8.5$eV.}
\label{fig:errorIdf}
\end{figure}

\begin{figure}[thb]
\vskip -0.5 cm
\hspace*{-01.05cm}\includegraphics[width=0.6\textwidth]{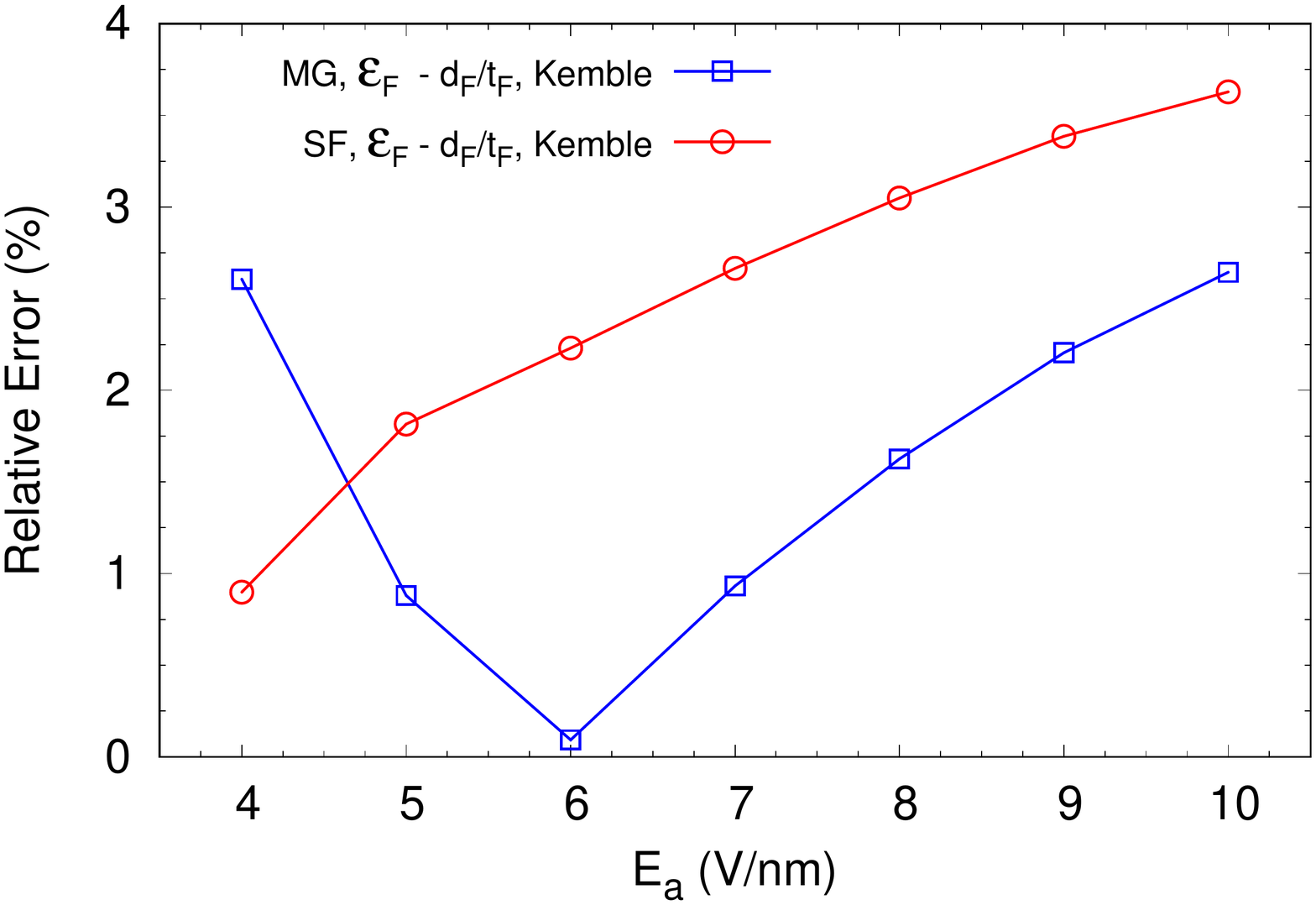}
\vskip -0.75 cm
\caption{As in Fig.~\ref{fig:errorIdf} using the correction term in $J_{\mg}^m$ and $J_{\sh}^m$.
The errors are seen to reduce at higher fields in both cases.}
\label{fig:errorIdfK}
\end{figure}

The linearization at $\ce_F$ results in large errors at higher fields despite the
use of the correction factor arising from the Kemble form. Figures \ref{fig:errorIdf}  and
\ref{fig:errorIdfK} show the relative errors in $J_{\mg}^m$ and $J_{\sh}^m$,
without and with the correction factor respectively.
In both figures, $\ce_m$ corresponds to the energy where the peak
of the normal energy distribution occurs. 
This value is closely approximated by
$\ce_m \approx \ce_F - d_F/t_F$ where $d_F = g\sqrt{\phi}/E_l$ and $t_F = 1 + y_F^2/9 - (y_F^2/9)\ln(y_F)$
with $y_F = c_S \sqrt{E_l}/\phi$. For the SF results, we have also numerically
determined the location of the maximum and found little change in the errors.

The errors in both MG and SF reduce compared to the results for $\ce = \ce_F$.
The MG case undergoes a transition from under-prediction to over-prediction
(compared to the benchmark) as the field increases while the SF method consistently over-predicts
the net current for all values of $E_a$. As in case
of the expansion at $\ce_F$, the MG approximation gives better results which improves further
on use of the correction factor due to the Kemble form of transmission coefficient (see
Fig. \ref{fig:errorIdfK}). In summary, the use of $J_{\mg}^m$  with
$\ce_m = \ce_F - d_F/t_F$ is found to have errors within 3\% of the exact WKB result over a wide
range of fields.

\section{Discussions and Conclusions}

The shape-factor $\sigma$, even with the quadratic approximation, is found to give very good results
for all local fields, so long as the integration over the electron energies is carried out numerically.
The error is found to be below 1\% compared to the WKB result using the exact Gamow factor,
and decreases at higher fields. When the transmission coefficient is determined using the
Forbes approximation for the barrier form correction factor, the errors are much larger
but improves at higher fields. In contrast, a linearization of the Gamow factor
using the  quadratic approximation for the shape-factor, leads to errors that are somewhat larger compared
to the Murphy-Good-Forbes approach.  Thus, if a reasonably accurate
analytic form for the cold field emission
current density is required, the  current density $J_{\mg}^m$ as in Eq.~(\ref{eq:MGshift})
with $\ce_m = \ce_F - d_F/t_F$, corresponding approximately to the peak of the normal energy distribution,
is suitable for its accuracy and relative ease of use. The ease of use can be further improved
by ignoring the correction factor  provided the fields involved are not
too high. Note that the errors on using $J_{\text{SF}}^m$ with $\ce_m =  \ce_F - d_F/t_F$, is
only marginally higher and hence there is very little to choose between the two 
except for a simpler expression for $J_\mg^m$.

It is possible that the average error may reduce further on choosing another point
of linearization or improving upon the algebraic approximations for $\nu(y)$ and
$\sigma(y)$ considered here. Such improvements would obviously lead to more involved expressions
for the current density making them more cumbersome to use for analyzing experimental data.
It must also be noted that the benchmark chosen here
is the exact-WKB method and errors may be quite different, and even large
compared to the errors presented here, if the exact current is instead chosen for comparison.
Such errors are found to follow an approximate trend and can be minimized
(though not eliminated) by using a correction factor dependent on $\ce_F/\phi$
(as shown in Ref~[\onlinecite{db_rr_2021}]) along with $J_{\mg}^m$.

Finally, since a shifted point of linearization and the use of the Kemble correction are
found to be important in reducing errors, the notation $J_\mg^m$ may be reserved for the current density
given by Eqns. (\ref{eq:MGshift}) and (\ref{eq:BFNshift}) for $\ce_m \neq \ce_F$, in order to
distinguish it from the standard MGCD denoted by $J_\mg$, which corresponds to $\ce_m = \ce_F$ and 
does not have any Kemble-correction factor.

\section{Acknowledgements} The author acknowledges discussions with Rajasree Ramachandran and Raghwendra Kumar.

\section{Author Declarations}

\subsection{Conflict of interest} There is no conflict of interest to disclose.
\subsection{Data Availability} The data that supports the findings of this study are available within the article.



\begin{thebibliography}{99}
\bibitem{TC} In quantum mechanics, the transmission coefficient is the ratio of the current density transmitted through a barrier and the current density incident on the barrier. It is also referred to as the transmission probability.
\end{thebibliography}
\end{document}